# RelSifter: Scoring Triples from Type-like Relations

## The Samphire Triple Scorer at WSDM Cup 2017

Prashant Shiralkar[1,†], Mihai Avram[1], Giovanni Luca Ciampaglia[2]
Filippo Menczer[1,2], Alessandro Flammini[1,2]
[1]School of Informatics and Computing
[2]Network Science Institute
Indiana University, Bloomington, IN USA

## ABSTRACT

We present RelSifter, a supervised learning approach to the problem of assigning relevance scores to triples expressing type-like relations such as 'profession' and 'nationality.' To provide additional contextual information about individuals and relations we supplement the data provided as part of the WSDM 2017 Triple Score contest with Wikidata and DBpedia, two large-scale knowledge graphs (KG). Our hypothesis is that any type relation, i.e., a specific profession like 'actor' or 'scientist,' can be described by the set of typical "activities" of people known to have that type relation. For example, actors are known to star in movies, and scientists are known for their academic affiliations. In a KG, this information is to be found on a properly defined subset of the second-degree neighbors of the type relation. This form of local information can be used as part of a learning algorithm to predict relevance scores for new, unseen triples. When scoring 'profession' and 'nationality' triples our experiments based on this approach result in an accuracy equal to 73% and 78%, respectively. These performance metrics are roughly equivalent or only slightly below the state of the art prior to the present contest. This suggests that our approach can be effective for evaluating facts, despite the skewness in the number of facts per individual mined from KGs.

## 1. INTRODUCTION

In RDF, *triples* take the form `(subject, predicate, object)`, where the `subject` and `object` are entities, and `predicate` is a relation. The task of the triple scoring challenge at the WSDM Cup 2017 [2, 8] was formulated as follows:

> Given a triple from a type-like relation, compute a score (in the range 0 through 7) that measures the relevance of the statement expressed by the triple compared to other triples from the same relation.

Here, a triple with score of 7 is considered as the most relevant, whereas that with a score of 7 is the least relevant. The triple scoring problem was first introduced by Bast et al. [1], who also present a thorough review of the relevant literature.

Because the goal is to compute a numeric score for RDF triples, the task can be seen as a special case of fact-checking relational statements [6]. In this paper, we often use the term "fact" to refer to such triples. However, there is distinction between the two problems. The goal of fact checking is to discern true, and possibly unknown facts, from dubious ones. In the present task, instead, all facts are true, but some are less relevant than others. So in some sense, relevance scoring can be seen as a more subtle form of fact checking.

†Corresponding author. Email: pshiralk@indiana.edu

In the specific context of the WSDM Cup 2017, a special class of triples was considered: triples expressing a so-called *type-like relation*, or TLR. These triples are important because through their predicate and object they specify information about the type of the subject entity, and they often occur as results of search engine queries. Thus, given a subject, knowing how important, or *relevant*, a particular object is (or viceversa) is a critical problem in information retrieval.

Two particular TLRs were used in the WSDM Cup 2017 task: *professions* and *nationalities*. Consequently, subjects were people in all cases, and objects were specific professions or nationalities. An example of two such triples from the training dataset is `(Wolfgang Amadeus Mozart has-profession Composer)` and `(Wolfgang Amadeus Mozart has-profession Pianist)`. Though both triples are true, the fact that Mozart was a composer is intuitively more relevant than the fact that he was a pianist. The relevance scores associated to type-like triples are often referred to as *triple scores*.

In our approach, we mainly rely on a set of structured facts about persons and their professions or nationalities available in two large-scale knowledge bases: DBpedia [4] and Wikidata [7]. The rationale for employing these two knowledge bases is their rich, cross-domain information about many real-world entities and associated relations. A knowledge base can be naturally seen as a graph by considering its triples as a list of edges with additional semantics given by edge labels — the predicates. In the following, we motivate the intuition behind our approach using the 'profession' TLR; however, the same intuition is applicable to any TLR.

Our idea is simple. We characterize each profession object, like 'actor' or 'scientist,' by the set of its most pertinent "activities" — the facts that typically denote people known to have that profession. These are simply other predicates of the KG. We first identify the set of all entities `s` of the KG which are the subject of some triple `(s, has-profession, actor)` — the set of all actors. We define the set of all labels of outgoing edges of nodes in this set, that is, all the predicates of any triple in which an actor is a subject entity. Of course, this set of predicates will contain several predicates that are not informative of being an actor. A good example would be the predicate `bornIn` — everybody is born somewhere after all. Some other predicates, however, may be pertinent to being an actor. For example, actors are known to star in movies or plays.

We approach the problem of identifying the most pertinent activities for a profession by means of TF-IDF, and pick only the top $k$ predicates as the *activities* associated to having profession $p$. A pooled set of these top $k$ activities for all the professions forms the feature set for our learning problem. For each triple in the provided training set, we then measure the extent of "overlap" between ob-served activities of the person in the triple and the defining activities for the object as identified earlier. That is, we create a sparse feature vector for each triple, where a non-zero entry is equal to the TF-IDF value for the activity in which the `subject` is engaged and overlaps with the defining activities of the `object`.

We first review in Section 2 all the data sets and pre-processing steps carried out to reconcile the data provided in the challenge and those in the knowledge graphs. In Section 3, we discuss a TF-IDF approach for characterizing the most informative activities. Section 4 describes how we use the profession and nationality representations in learning a model to predict triple scores. Finally, we present a summary of our results and a few concluding remarks.

## 2. DATA

The given data consisted of a knowledge base of 499,244 `(person, profession)` and 318,779 `(person, nationality)` pairs that featured 200 professions, 100 nationalities, and 385,426 persons. We supplemented this data with two large knowledge graphs, namely DBpedia and Wikidata. Table 1 gives a summary of these undirected KGs we used in this work.

**Table 1: Statistics of Knowledge Graphs.**

| KG | Nodes | Relations | Triples |
|---|---|---|---|
| DBpedia | 6.1M | 663 | 48M |
| Wikidata | 29.4M | 839 | 234M |

### 2.1 DBpedia

DBpedia is a knowledge base of structured facts extracted primarily from infoboxes of Wikipedia articles. We used its ontology, instance-types, and mapping-based properties information from the April 2016 RDF dump[1] to form our knowledge graph. Most of the person, profession and nationality information could be readily reconciled with entities in DBpedia after replacing spaces by underscores. For a small fraction that could not be resolved, we performed entity resolution in a brute-force manner using the redirect information provided by DBpedia. Ultimately, in DBpedia, we could not match 38,306 (person,profession) pairs out of 499,244 (8%), and 8,483 (person,nationality) pairs out of 318,779 (3%).

### 2.2 Wikidata

Wikidata follows a richer data model than DBpedia. To harmonize the two data sources we used the dataset of *simple statements*, which excludes all contextual and provenance information. We downloaded the taxonomy, class hierarchy of Wikidata properties, class membership information, and simple statements as an RDF dump from August 2016.[2] Since entities and properties in Wikidata are identified by an opaque item identifier (e.g. `Q42` refers to `Douglas Adams`, while `P26` refers to the `spouse` relation), we also downloaded terms information to get their corresponding labels. To perform entity resolution and/or de-duplication in some cases, we used the site-link information that provides links from Wikidata to Wikipedia articles. Our experience, while working on this problem, suggests that Wikidata has comparatively more information about the entities than DBpedia. This is also evident from Fig. 1. Nevertheless, we report results from both knowledge graphs.

[1]wiki.dbpedia.org/downloads-2016-04

[2]tools.wmflabs.org/wikidata-exports/rdf/exports/20160801/dump_download.html

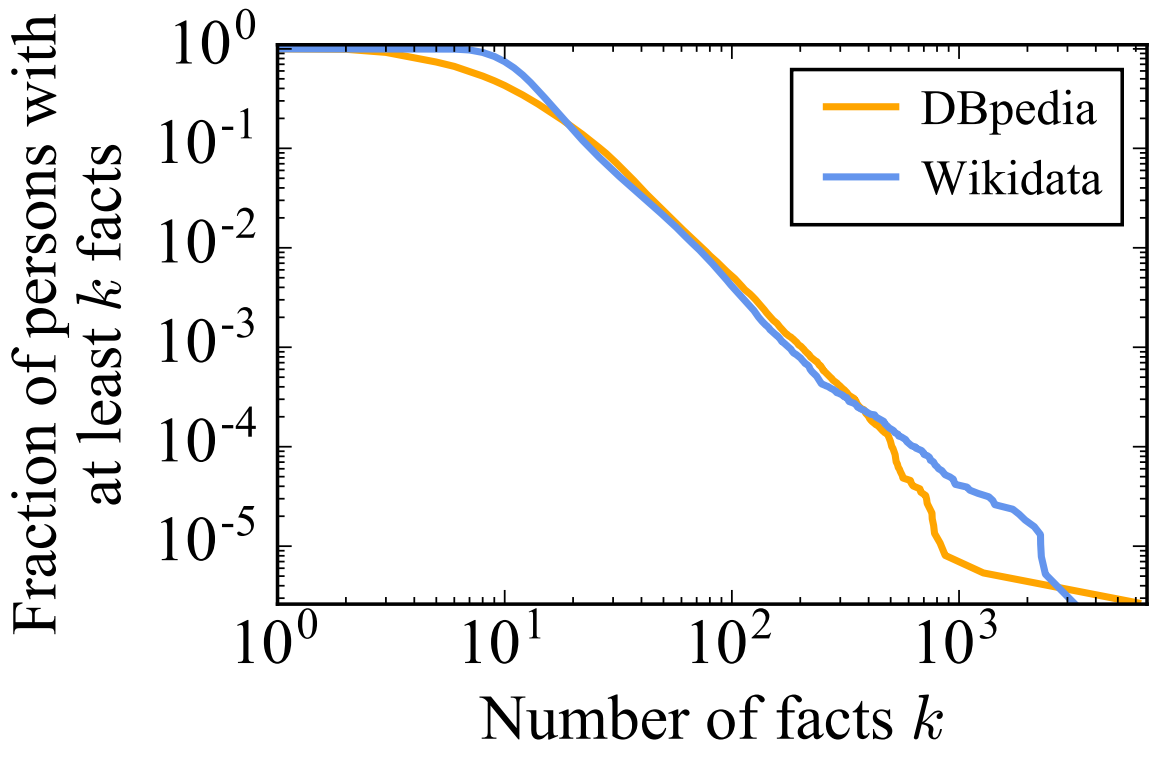

**Figure 1: Complementary cumulative distribution of the number of facts per person in the two datasets.**

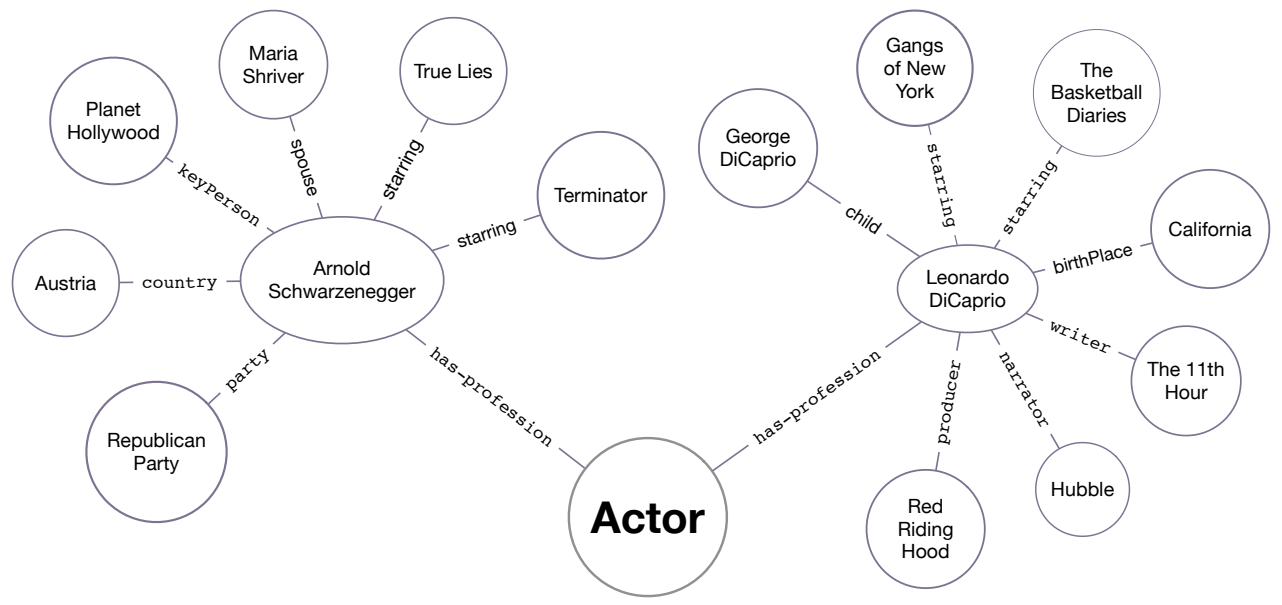

**Figure 2: In this example from DBpedia, `starring` is an activity pertinent to the profession `Actor`.**

### 2.3 Wikipedia Abstracts

Because our set of people was drawn from Wikipedia, we also considered the text of their Wikipedia entries available in the dataset of short and long abstracts published as part of DBpedia. We chose to use abstracts as opposed to complete Wikipedia article text after making a crude observation that most information describing a person, profession or nationality is usually available as part of the initial text, which forms the abstract of the article. For entities that did not have a long abstract, we settled with a short abstract.

## 3. FEATURE CONSTRUCTION

### 3.1 TF-IDF Features

We measure the pertinence of an activity to a TLR object $o$ by combining two dimensions of its informativeness: popularity and focus. Popularity of an activity aims to capture the intuition that it is representative of an object profession (or nationality) if it is popular among people known to have that profession (or nationality). Activity information is found on a properly defined subset of the second-degree neighbors of the type relation (Fig. 2). E.g., one of the top activities for profession `Actor`, with this metric, is `starring`.

Formally, let $U$ represent the universe of all people in the KG, $S_o$ the set of people with profession $o$, and $R_o$ the set of relations associated with people in $S_o$. Let us consider a predicate $r \in R_o$ and denote by $g_{S_o}(r)$ the number of person entities in $S_o$ with activity $r$, and by $g_U(r)$ the number of person entities in $U$ with

**Table 2: Top 5 activities for a few professions in Wikidata per combined pertinence.**

| Rank | Pianist | Lyricist | Architect | Talk show host | Mathematician |
|---|---|---|---|---|---|
| 1. | composer | lyrics by | architect | presenter | proved by |
| 2. | follower of | score by | architectural style | production company | solved by |
| 3. | instrument | librettist | structural engineer | narrator | doctoral student |
| 4. | lyrics by | performer | main building contractor | executive producer | doctoral advisor |
| 5. | record label | composer | notable work | influenced by | field of work |

**Table 3: Top 5 activities for a few nationalities in Wikidata per combined pertinence.**

| Rank | United States of America | Italy | Australia | India | Soviet Union |
|---|---|---|---|---|---|
| 1. | defender | production designer | spoken text audio | pronunciation audio | backup or reserve team or crew |
| 2. | military casualty classif. | feed URL | solved by | solved by | astronaut mission |
| 3. | executive producer | party chief repr. | cuisine | audio | academic degree |
| 4. | investor | executive body | handedness | filmography | corporate officer |
| 5. | filmography | powerplant | bowling style | bowling style | contributing factor of |

activity $r$. The pertinence of $r$ to $o$ by popularity $P$ is:

$$P_o(r) = \log(1 + g_{S_o}(r)) \cdot \log\left(\frac{|U|}{g_U(r)}\right).$$

Another form of pertinence of an activity is its role in defining the 'focus' of a person. An activity may be representative of an object profession (or nationality) if it is relatively more frequent among people having that profession (or nationality) than among those who do not. For example, the top activity for `Actor` according to this metric is `cast member`, which is different from the top activity by popularity. Let us denote by $f_{S_o}(r)$ the frequency of $r$ for persons in $S_o$. Then the pertinence of $r$ to $o$ by focus $F$ is defined as:

$$F_o(r) = \log(1 + f_{S_o}(r)) \cdot \log\left(\frac{\sum_{r' \in R_o}(1 + f_U(r') - f_{S_o}(r'))}{1 + f_U(r) - f_{S_o}(r)}\right)$$

where the second term reflects the inverse relative frequency of relations in persons *not* having the profession $o$. The 1 inside log in above formulae ensures that the term is well-defined.

We finally combine these measures of pertinence by popularity and focus to define the *combined pertinence* $C$ as the product

$$\mathrm{C}_o(r) = \mathrm{P}_o(r) \cdot \mathrm{F}_o(r).$$

We used combined pertinence to identify the most informative activities of a TLR. Tables 2 and 3 show a few examples of the top $k$ activities, for $k = 5$, in decreasing order of their combined pertinence. The top $k$ activities of all TLR professions (or nationalities) when pooled together forms the set of features for our learning algorithm. As one might expect, the set of activities of related professions do overlap. For example, the activity `lyrics by` is among the top $k$ activities of both `Pianist` and `Lyricist`. However, its position differs in the two rankings.

## 3.2 Text Features

As an alternative approach, we characterized the object profession (or nationality) of a person based on the abstract of the corresponding Wikipedia entry. We pre-processed the information by removing stop-words and performing lemmatization. To perform stemming, we used WordNetLemmatizer from the Python NLTK package [3]. By following the classic TF-IDF approach in the information retrieval literature, we extracted the top-$k$ words as features.

# 4. TRIPLE SCORE LEARNING

Given a subject-predicate-object triple, where the subject is a person entity and the object a profession or nationality, we wish to learn a model that predicts the relevance score in a discrete, ordinal range between 0 and 7. In this section we describe how we solve this relevance scoring problem using a supervised learning algorithm.

Our idea is to measure the overlap between the activities that describe the subject entity and those characterizing the object profession or nationality.

The training set provided by the WSDM Cup task consists of 515 profession triples and 162 nationality triples. We construct a feature vector for each of these triples in the following way: if there is a triple in the KG whose subject corresponds to the subject of the triple being tested, and the predicate corresponds to any activity of the TLR being tested, then the corresponding entry of the feature vector for the triple being tested takes the range-normalized combined pertinence value of the activity, and zero otherwise. The resultant sparse feature matrix forms our input for the learning algorithms.

To take advantage of the ordinal structure in the labels, we considered Ordinal Logistic Regression (OLR) [11], which is a generalization of logistic regression for ordinal data. Since all mistakes are not equal in the case of ordinal data, we chose the *all-thesholds loss* that penalizes predictions made farther away from the expected output [11]. We also considered two other learners, namely Random Forest (RF) and Adaboost with decision trees as base learners, by treating the problem as a multi-class classification.

To control over-fitting we performed 10-fold cross-validation. For OLR, we experimented with the following penalty values for $\alpha$: 1, 5, 10, 15, 20, 50, 75, 100, 250, 500, 1000, whereas for RF and AdaBoost we tried bags of 10, 50, 100, 250 and 500 base estimators. For every algorithm, we selected the best model by a grid search over the range of their respective parameters.

All our code has been implemented in Python 2.7 and is available at github.com/tira-io/samphire. We used `scipy.sparse` matrices to represent the knowledge graphs. For feature extraction, training models and making predictions, we used the `sklearn` package. For OLR, we used a package called `mord` [9] that implements the approaches in [11] following the same API as `sklearn`.

# 5. EVALUATION RESULTS

As mentioned earlier, we experimented with two KGs: DBpedia and Wikidata. For each of the two KGs, we extracted top-$k$ activities for every object profession and nationality by their combined pertinence, built a feature matrix corresponding to these features, and evaluated performance using the three learning algorithms. For any triples whose entities could not be resolved during pre-processing, we chose to make random predictions. Fig. 3 shows the results obtained by experimenting with different values of $k$. Here, we only report on accuracy (as defined for the task) as our performance measure. For details on the Average Score Difference and Kendall Tau for our deployed (best) models, we refer the reader to the overview of the WSDM Cup 2017 [2, 8]. For both KGs, Random Forest worked better in case of profession, whereas Ordinal Logistic Regression was the outright winner for the nationality relation. In general, Adaboost delivered relatively poor performance.

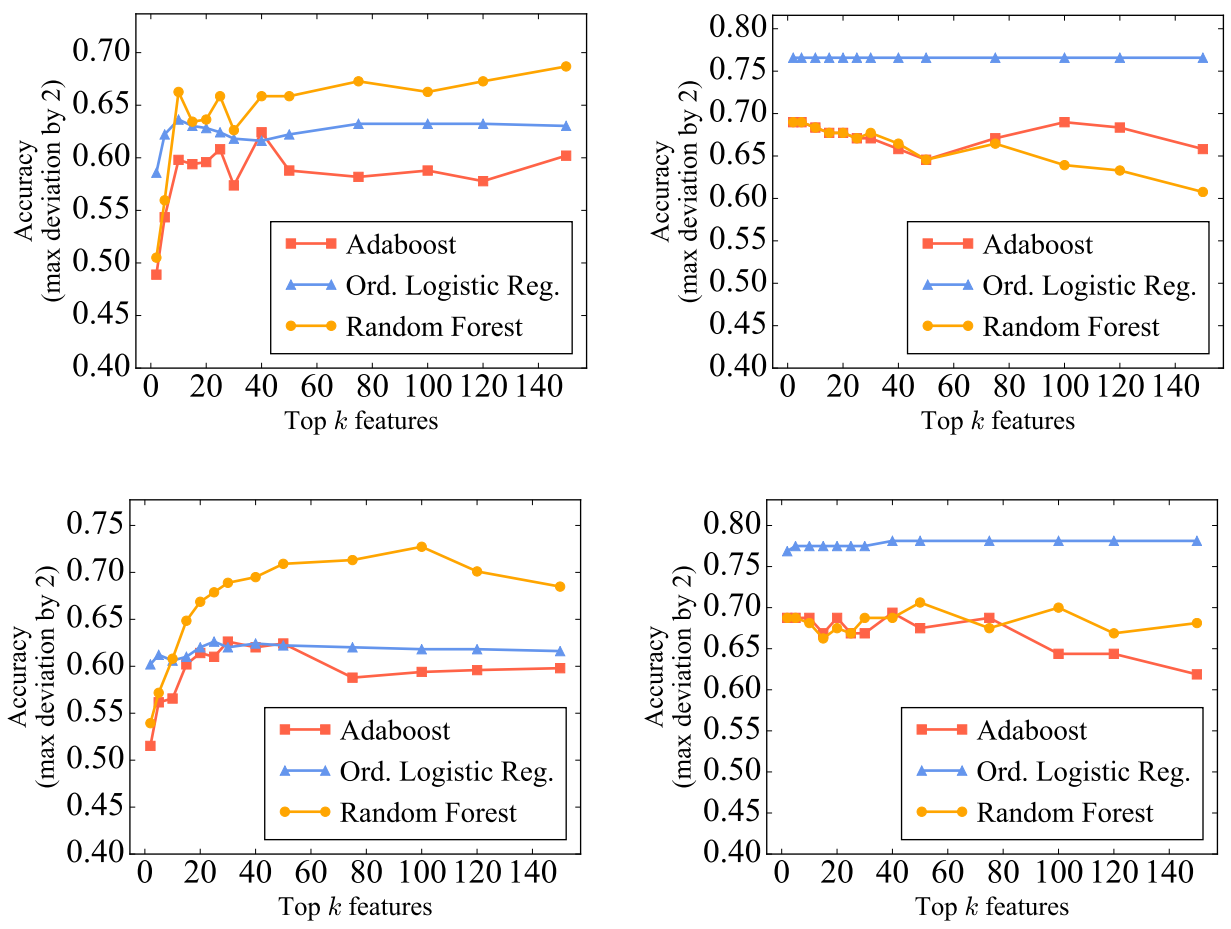

**Figure 3: Performance by combined pertinence. Top: DBpedia; Bottom: Wikidata. Left: Profession; Right: Nationality.**

The performance in both tasks is roughly equivalent or lags by a small margin behind the best performing algorithms of the prior state of the art [1], which we considered as our baseline. Our best models resulted in 73% accuracy for the profession relation and 78% for nationality relation. It is worth noting that the simplicity of our approach lies in having only a single model for a given TLR, as opposed to one for every profession (or nationality).

To raise performance further, we also tried leveraging information in the textual content of abstracts for each of the entities. We computed activities based on abstracts for both nationality and profession, constructed feature matrices, and built models for a few values of $k$. The performance from this text-based approach was not any better than the KG-based models.

Having two different models — one from the KG and another based on the text of abstracts — we also experimented with *stacking* models [5], and with building models by combining features from the two approaches.

Although we did not try all combinations of $k$ and parameter values, a few experiments did not result in significantly boosted performance as we initially hoped. Hence, we exclude results from the text-based and from the stacking-based models. Since models built on features engineered from Wikidata delivered the best performance, we deployed those in our final submission on the TIRA platform [10].

## 6. CONCLUSION

We addressed the problem of assigning relevance scores to type-like triples by following a supervised learning approach and leveraging facts in large knowledge graphs such as Wikidata and DBpedia. We found that the performance of our models was almost on par with the state of the art prior to the task [1]. This demonstrates that our approach is effective for the problem even with minimal amount of labeled training information. However, the significant difference between human judged scores and current model performance indicates that ample room for improvement remains.

Our experience suggests that besides skewness in the number of facts per individual, there is also skewness in the number of relations per TLR object, which we believe led to a suboptimal choice of the activities. Although the current state of KGs has shown great value, a lot more can be done on the relation extraction side, since that continues to be a major bottleneck for assembling more structured facts about these entities.

In our work, we only used local information about an individual to measure the overlap of his/her activities with that of the TLR object. In future work one could build upon this idea and include larger neighborhoods of the individual to take advantage of greater context available in the KG. In summary, we conclude that the problem is challenging and warrants more research.